\documentclass[prd,twocolumn]{revtex4-2}
\usepackage{graphicx}
\usepackage{amssymb}
\usepackage{color}
\usepackage{enumerate}
\usepackage{amsmath}

\newcommand{\be}{\begin{equation}}
\newcommand{\ee}{\end{equation}}
\newcommand{\bea}{\begin{eqnarray}}
\newcommand{\eea}{\end{eqnarray}}

\begin{document}

	\title{Condensation temperature of strongly interacting $^{39}K$ condensates in the mean-field and semi-classical approximations.}

	\author{Fabio Briscese}\email{fabio.briscese@uniroma3.it, briscese.phys@gmail.com}
	
	\affiliation{Università Roma Tre, Dipartimento di Architettura, via Madonna dei Monti 40 00184 Rome, Italy,}
	\affiliation{ Istituto Nazionale di Alta Matematica Francesco		Severi, Gruppo Nazionale di Fisica Matematica, Citt\`{a} Universitaria, P.le A. Moro 5, 00185 Rome, Italy.}

	\begin{abstract}

We consider the effect of inter-atom interactions on the condensation temperature $T_c$ of an atomic Bose-Einstein condensate. We find an analytic expression of the shift in $T_c$ induced by interactions with respect the ideal non-interacting case, in the mean-field and semi-classical  approximations. Such a shift is expressed  in terms of the ratio $a/\lambda_{T_c}$ between the s-wave scattering length $a$ and the thermal wavelength $\lambda_{T_c}$. 
This result is used to discuss the tension between mean-field predictions and  observations in strongly interacting $^{39}K$ condensates. It is shown that such a tension is solved taking into account the details of the Feshbach resonance used to tune $a$ in the experiments.

	\end{abstract}

\maketitle


The effect of inter-atom interactions in atomic Bose-Einstein condensates (BECs) in the mean-field approximation and beyond (see \cite{dalfovo,ps} for review) has been extensively studied, from the first works of Lee, Huang, and Yang \cite{a1,a2} to more recent years \cite{a3,a4,a5,a6,a7,a8,a9,a10,a11,a12,a13,a14,a15,c1,c2,c3,c4,c5,c6,c7,c8}.  The mean-field approximation accounts for leading inter-particle interactions, which are crucial for the thermal equilibrium in the condensate, while beyond mean-field effects are due to quantum corrections. 

In this paper we focus on the dependence of the condensation temperature $T_c$ on the scattering length in the mean-field approximation.
In the case of uniform BECs,
interactions are irrelevant in the mean-field approximation \cite{a7,a10,a11,a12}. However,  beyond-MF effects related to quantum correlations between bosons near the critical point produce a shift in the condensation temperature with respect to the ideal non interacting case given by \cite{a7,a8}

\begin{equation}
\Delta T_{c}/T_{c}^{0}\simeq \, 1.8 \, (a/\lambda_{T^0_c})
\label{deltaTsuTUNIFORM}
\end{equation}
with $\Delta T_{c}\equiv T_{c}-T_{c}^{0}$, where  $T_{c}$ is the critical temperature of the gas of interacting bosons, 

\be
T_{c}^{0}= \left(n/\zeta(3/2)\right)^{2/3}(2\pi\hbar^2/m k_B)
\ee 
the condensation temperature in the ideal non-interacting case, 
\be\label{definition lambda T}
\lambda_{T_c}\equiv \sqrt{2 \pi \hbar^2/(m k_B T)}
\ee 
the
thermal wavelength at temperature $T$, $m$ the boson mass, $n$ the boson number
density, $a$  the s-wave scattering length used to parameterize
inter-particle interactions, and $k_B$ the Boltzmann constant \cite{dalfovo,ps}.

However, laboratory atomic condensates are confined in magnetic  and optical traps, indeed they are not uniform. Trapped condensates can be described as a
system of $N$ bosons in an external 
harmonic potential. 
Neglecting the effect of inter-atom interactions,  the condensation temperature
is \cite{dalfovo,ps}
\be\label{T_C^0}
k_B T^0_c = \hbar \omega \left(N/\zeta(3)\right)^{1/3}\, ,
\ee
while, when such an effect is taken into account, it causes a shift in $T_c$ given by

\begin{equation}
\frac{\Delta T_c}{T^0_c}\simeq b_1 (a/\lambda_{T^0_c}) + O(a/\lambda_{T^0_c})^2 \, . \label{deltaTsuNONTUNIFORM1}
\end{equation}
The coefficient $b_1$ can be estimated   in the mean-field and semi-classical approximations, giving $b_1 \simeq -3.426\ldots$ \cite{dalfovo}. This value is in excellent agreement with
laboratory measurements of $\Delta T_{c}/T_{c}^{0}$
\cite{b1,b2,b3,condensatePRL} for sufficiently weak atom-atom interactions, i.e.  small
$a/\lambda_{T^0_c}\ll 0.01$.

High precision measurements of the condensation temperature of $^{39}K$ in the strongly interacting regime have detected $O(a/\lambda_{T^0_c})^2$
effects in $\Delta T_{c}/T_{c}^{0}$, see \cite{condensatePRL}. This has been achieved exploring the range $5\times 10^{-3} < a/\lambda_{T^0_c} < 4 \times
10^{-2}$ for parameters $N\simeq (2-8) \times 10^5$, $\omega
\simeq (75-85) Hz$,   $T_c \simeq (180-330) nK$, and $\lambda_{T_c}\simeq (4.8-6.5)\times 10^{-7}m$. The measured temperature shift  is fitted by a second order polynomial

\begin{equation}
\frac{\Delta T_c}{T^0_c}\simeq b_1 (a/\lambda_{T^0_c}) + b_2 (a/\lambda_{T^0_c})^2 \, , \label{deltaTsuTsecondorder}
\end{equation}
with  $b_1
\simeq -3.5 \pm 0.3$ and $b_2 \simeq 46 \pm 5$ \cite{condensatePRL}. Theoretical estimations of $b_2$ in the mean-field and semi-classical approximations strongly disagree with its best-fit value. For instance,
a value $b_2\simeq 11.7$ has been obtained in \cite{condensatePRL2} numerically, which is  excluded by data at $\sim 6\,\sigma$ level. 

More in general, it has been shown \cite{briscese2}  that, in the mean-field and semi-classical approximations, one has

\begin{equation}
\frac{\Delta T_c}{T^0_c} \approx b_1 (a/\lambda_{T^0_c}) + b_2
(a/\lambda_{T^0_c})^2 +  \psi\left[a/\lambda_{T^0_c}\right] \, ,
\label{deltaTsuT Asymptotic}
\end{equation}
where  $\psi\left[z\right]$ is  non-analytic in $z=0$, with $\psi^\prime\left[0\right]=\psi^{\prime\prime}\left[0\right]=0$, while its higher derivatives are divergent in $0$. The explicit form of the function $\psi[z]$ is not known, while the parameters $b_1$ and $b_2$  can be estimated analytically, giving $b_1 \simeq -3.426\ldots$ and $b_2 \simeq 18.776\ldots$, with $b_2$ still in disagreement with \cite{condensatePRL} at $\sim 4\, \sigma$ level. 

We mention that a different estimation of $\Delta T_{c}/T_{c}^{0}$ can be  obtained by means of lattice simulations \cite{c1}, giving 
\begin{equation}
	\label{deltatsut log}
	\frac{\Delta T_c}{T^0_c} \simeq c_1 (a/\lambda_{T^0_c}) + \left(
	c'_2 + c''_2 \ln[a/\lambda_{T^0_c}]\right) (a/\lambda_{T^0_c})^2 \, ,
\end{equation}
with $c_1 \simeq -3.426$, $c'_2 \simeq -32
\pi \zeta[2]/3\zeta[3] \simeq -45.9$ and $c''_2 \simeq -155$, that also disagrees with the measurements reported in
\cite{condensatePRL}.

In what follows, we begin finding  an analytic expression of $\Delta T_{c}/T_{c}^{0}$ as a function of the ratio $a/\lambda_{T_c}$ in the mean-field and semi-classical approximations. This result confirms the tension between mean-field predictions  and measurements in strongly-interacting $^{39}K$ condensates \cite{condensatePRL}. We then show that such a tension is due to  an overestimation of the s-wave scattering length, owing to a standard description of the Feshbach resonance used to tune $a$ in \cite{condensatePRL}, which is inaccurate in the strong interacting regime. We then argue that, when the correct expression of $a$ in term of the Feshbach magnetic field $B$ is considered, and the measurements of the temperature shift are rescaled accordingly, the data in \cite{condensatePRL} show perfect match with mean-field predictions.

Let us start considering a
system of $N$ bosons trapped in an external 
harmonic isotropic potential \mbox{$V(x) = m \omega^2 x^2/2$} (the generalization to anisotropic traps and other potentials is straightforward), with $\omega$ the frequency of the trap, and $m$ the mass of the bosons.
In the mean-field approximation, the BEC is described as a system of bosons that experience a mean-field interaction potential 
$\propto a \, n$ where $a$ is the scattering length and $n$ is the density of
bosons, accounting for atom-atom interactions. Therefore,  the single-particle energy in phase space is given by \cite{dalfovo,ps}

\begin{equation}\label{energy interaction}
E =  \epsilon(p,x) + 2 g \, n(x,g)
\end{equation}
where 
\be\label{definition g}
g = \frac{4 \pi \hbar^2}{m} \,a \, .
\ee 
Moreover, in the semi-classical limit one has $\epsilon(p,x) \equiv p^2/2m + V(x)$, see
\cite{dalfovo,ps} for details. Under these assumptions, the number of bosons in
the excited thermal spectrum is given by

\begin{equation}\label{semiclassicalNth1}
N_{th}= \int \frac{d^3x d^3p}{(2\pi \hbar)^3}
\left(\exp\left[\frac{E-\mu}{k_B T}\right]-1\right)^{-1},
\end{equation}
where  $\mu$ is the chemical potential. This expression can be rearranged in the form $N_{th} = \int d^3x\,n (x,g)$, where the number density of the bosons in the thermal cloud is given by

\begin{equation}
\begin{array}{ll}
\label{semiclassicalNth2}
n_{th}(x,g)=  \int \frac{ d^3p}{(2\pi \hbar)^3}
\left(\exp\left[\frac{E-\mu}{k_B T}\right]-1\right)^{-1}=\\
\\
\lambda_{T_c}^{-3}\, g_{3/2}\left[\exp\left[ - \frac{V(x)+2
	g \, n(x,g)-\mu }{k_B T}\right] \right]\, ,
\end{array}
\end{equation}
where $\lambda_{T_c} = \sqrt{2 \pi \hbar^2 / (m k_B T)}$ and
$g_{\alpha}[z] = \sum_{k=1}^\infty z^k/k^{\alpha}$ is the
Polylogarithmic or Boltzmann function of index $\alpha$, see \cite{dalfovo,briscese2} for details. This is a
consistency relation which can be used, at least in principle, to extract
$n(x,g)$. Nevertheless, we will show that for our purposes we do not need to find $n(x,g)$ explicitly.

When the system reaches the condensation
temperature $T_c$, the chemical potential $\mu$  equals  the energy of the ground state, corresponding to the minimum of $E$. It can be shown \cite{dalfovo,briscese2} that such a minimum corresponds to

\begin{equation} \label{numericalmuc}
\mu_c = 2 \, g \, n(0,g) .
\end{equation}
Moreover, at the temperature  $T_c$ the number of bosons in the condensate $N_0$ is still zero. Thus from
(\ref{semiclassicalNth1}-\ref{numericalmuc}) one has

\begin{equation}\label{semiclassicalNth3}
N= \int \frac{d^3x}{ \lambda_{T_c}^3}g_{3/2}\left[\exp\left[ - \frac{V(x)+2
	g \left(n(x,g)-n(0,g) \right)  }{k_B T_c}\right] \right] \, ,
\end{equation}
and this expression must be inverted to give $T_c$ as a function of $g$, $N$ and $\omega$. 

In practice, extracting $T_c$ from (\ref{semiclassicalNth3}) is a difficult task; indeed (\ref{semiclassicalNth3}) is expanded around $g=0$ so that $T_c$ and its derivatives at $g=0$ are found, see \cite{dalfovo,briscese2}. This is realized at first order deriving both sides of (\ref{semiclassicalNth3}) with respect to $g$ and collecting the terms proportional to $\partial_g T_c$. Thus, $\partial_g T_c(g=0)$ is calculated setting $g=0$ in the integrals. Iterating this procedure at second order, one finds $\partial_g^2 T_c$ at $g=0$ and, proceeding in such a way, one can, at least in principle, express $T_c$ as an infinite Taylor series evaluated at $g=0$.
However, it is found that this procedure can not be iterated many times, since $g_{3/2}[z]$ is not analytic at $z=0$, as it has infinite first derivative there. Indeed, one has  divergent integrands at $x=0$ for $g=0$ in the integral expressions of higher derivatives of $T_c$. Therefore,  according to (\ref{deltaTsuT Asymptotic}), $\Delta T_c/T^0_c$ is determined at order $(a/\lambda_{T_c})^2$ up to an unknown function $\psi[a/\lambda_{T_c}]$ that is non-analytic at zero. As long as $\psi[z]$ remains unknown, one has no control of higher order contributions, even though they are expected to be small. Indeed, comparison of (\ref{deltaTsuT Asymptotic}) with data is not exhaustive.

Here we adopt a different strategy, which allows to find an analytic expression for $\Delta T_c/T^0_c$. We define the following variables:

\begin{equation}\label{variables definition}
\begin{array}{ll}
v\equiv \sqrt{\frac{V(x)}{k_B T_c}} =\sqrt{\frac{m \omega^2}{2k_B T_c}} \, x\\
\\
\Omega \equiv \frac{V(x)+2
	g \left(n(x,g)-n(0,g) \right)  }{k_B T_c}=v^2+\frac{2
	g \left(n(x,g)-n(0,g) \right)  }{k_B T_c} \,.
\end{array}
\end{equation}
Changing integration variable $x\rightarrow v$ in (\ref{semiclassicalNth3}), and using (\ref{definition lambda T}) one has

\begin{equation}\label{semiclassicalNth4}
N=\left(\frac{k_B T_c}{\hbar\omega}\right)^3 \frac{4}{\sqrt{\pi}} \int_0^\infty dv \, v^2\, g_{3/2}\left[e^{ -\Omega(v) }\right] \, .
\end{equation}
Note that, when $g=0$, one has $\Omega=v^2$ and the integral in the r.h.s. of (\ref{semiclassicalNth4}) gives $\sqrt{\pi}\zeta(3)/4$, so that one recovers (\ref{T_C^0}).  

In order to evaluate $T_c$, we should use the second equation in (\ref{variables definition}) to express $\Omega$ as a function of $v$ and $g$, and replace it in (\ref{semiclassicalNth4}). This is difficult to do in practice, but this issue can be overcome easily. Using (\ref{definition g},\ref{semiclassicalNth2})  one has

\begin{figure}	\begin{center}
		\includegraphics[height=5cm]{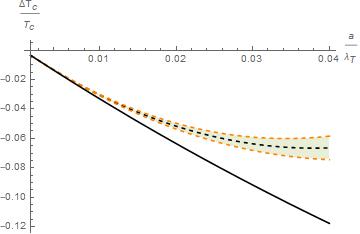}\qquad	\caption{We plot ${\Delta T_c}/{T^0_c}$ as a function of $a/\lambda_{T_c}$ in the range of values explored in \cite{condensatePRL}. The solid curve represents the mean-field result in (\ref{DT/T Final}),  the  dashed curve is the best fit (\ref{deltaTsuTsecondorder}), and the shadowed region is its $1\,\sigma$ contour.}
		\label{fig1}		
	\end{center}
\end{figure}

\be\label{definition Omega}
\Omega = v^2 + 4 \frac{a}{\lambda_{T_c}} \left(g_{3/2}\left[e^{-\Omega}\right]-\zeta\left(3/2\right)\right) \, ,
\ee
where we used $\Omega(x=0) = 0$ and $g_{3/2}\left[1\right]=\zeta\left(3/2\right)$.One can use  (\ref{definition Omega}) to express $v$ as a function of $\Omega$ and change integration variable in (\ref{semiclassicalNth4}). Indeed 

\be\label{definition v}
v(\Omega)=\sqrt{\Omega - 4 \frac{a}{\lambda_{T_c}} \left(g_{3/2}\left[e^{-\Omega}\right]-\zeta\left(3/2\right)\right)} \, ,
\ee
so that one has

\begin{equation}\label{semiclassicalNth6}
N=\left(\frac{k_B T_c}{\hbar\omega}\right)^3 \frac{4}{\sqrt{\pi}} \, I(a/\lambda_{T_c}) \, ,
\end{equation}
where

\begin{equation}\label{definition I}
I(a/\lambda_{T_c})\equiv  \int_0^\infty d\Omega \,\, v^\prime(\Omega) \, v^2(\Omega)\, g_{3/2}\left[e^{ -\Omega }\right] 
\end{equation}
with $v^\prime(\Omega) = \partial_\Omega v(\Omega)$. Finally, from (\ref{T_C^0}) and (\ref{semiclassicalNth6}) one has

\be\label{DT/T Final}
\frac{\Delta T_c}{T^0_c} \left(\frac{a}{\lambda_{T_c}}\right)= \left(\frac{I(0)}{I(a/\lambda_{T_c})}\right)^{1/3}-1\, ,
\ee
where $I(0)= \zeta[3]\sqrt{\pi}/4$.

Eq. (\ref{DT/T Final})  gives the analytic expression of the shift in the condensation temperature due to atom-atom interactions in the mean-field approximation in terms of the ratio $a/\lambda_{T_c}$. Such a function can be evaluated numerically, and it is plotted in fig.\ref{fig1} together with  the best fit curve (\ref{deltaTsuTsecondorder}). The plot shows clearly that the exact mean-field result (\ref{DT/T Final}) is far from closing the gap between the mean-field predictions and data. That means that the corrections due to the function $\psi[a/\lambda_{T_c}]$ in Eq.(\ref{deltaTsuT Asymptotic}) remain negligible in the range of values of $a$ explored in \cite{condensatePRL}, and the tension between mean-field predictions and data are not due to a poor evaluation of $T_c$.

In order to solve this problem, we have to analyze the details of the Feshbach resonance, which is used to tune the values of $a$ in the experiments. 
We remind that the s-wave scattering length is defined in terms of the two-body scattering amplitude $f(\theta)$ as
\begin{equation}\label{scattering length}
a\equiv - \lim\limits_{E\rightarrow 0} f(\theta) \end{equation}
where $E$ is the energy of the relative motion of the scattering particles, see e.g.  section 9 in \cite{ps} for details.  In correspondence of a Feshbach resonance, the energy of the scattering atoms, which have definite spin in the condensate cloud, becomes close to the energy $E_b < 0$ of a bound state in a different spin state. If there is a small coupling between these  states, this affects the value of the scattering length. What is more, the difference $E-E_b$  can be tuned changing the value of the external magnetic field $B$, as the scattering atoms and the bound state have different magnetic moment. It turns our that $a$ grows in magnitude when $B$ approaches a value $B_0$ of the magnetic field, where the resonance takes place. Moreover, the sign of the scattering length changes when $B$ crosses $B_0$, and the attractive/repulsive character of inter-atom interactions is swapped.

Therefore, in presence of a Feshbach resonance, the scattering atoms can decay into other channels,  so that the s-wave scattering length is replaced by a complex scattering length \cite{feschbach}

\begin{equation}\label{complex a}
\tilde{a}= a - i b = a_{bg} \left(1 + \frac{\Gamma}{-E_0+i \left(\gamma/2\right)}\right)
\end{equation}
where $E_0\equiv E_b+\delta E$, $\Gamma$ and $\delta E$ depend on the details of the interaction, $a_{bg}$ is the background value of $a$ far from the resonance, and $\gamma$ fixes the width of the resonance. 
Moreover, the thresholds for the elastic collision cross section $\sigma_{el}$
and the inelastic rate coefficient $K_{loss}$ are \cite{feschbach}
\be\label{kloss}
\sigma_{el}= 4 \pi (a^2+b^2)\,,\quad K_{loss} =\left( \frac{8 \pi \hbar}{m }\right) b \, .
\ee

As there is a difference  between the magnetic moment  $\mu_{atoms}$ of the separated atoms and the magnetic
moment $\mu_b$ of the bare bound state, the difference between the energy $E_b$ of the bound state   and the energy $E \rightarrow 0$ of the separated atoms  is $E_b = \delta\mu \left(B-B_b\right)$, where $B_b$ is the value of the magnetic field at which $E_b=0$, and $\delta\mu \equiv \mu_{atoms}-\mu_b$. That implies that the strength of inter-atoms interactions can be tuned varying $B$, indeed $\tilde{a}$ can be expressed as a function of $B$ as 

\begin{equation}\label{complex a 2}
	\tilde{a}= a_{bg} \left(1 - \frac{\Delta}{B-B_0 -i \Sigma} \right)   \,
\end{equation}
with 

\be
\Delta \equiv \frac{\Gamma}{\delta\mu} , \quad B_0 \equiv B_b-\frac{\delta E}{\delta\mu} , \quad \Sigma \equiv \frac{\gamma}{2\delta\mu}
\ee

\begin{figure}	\begin{center}
		\includegraphics[height=5cm]{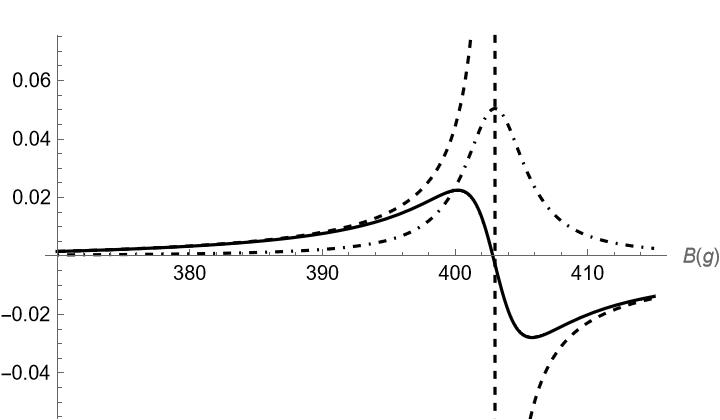}\qquad	\caption{We plot the functions $a/\lambda_{T_c}$ in (\ref{scattering length 2}), $a_p/\lambda_{T_c}$ in (\ref{scattering length 3}), and $b_p/\lambda_{T_c}$ in (\ref{collisional loss b})  against $B$ (dashed, solid, and dotted-dashed lines respectively) 
		for $\lambda_{T_c}\simeq 5.7 \times 10^{-7}m$, $\Delta \simeq -52 \, G$, $a_{bg}\simeq -29\, a_0$,  $B_0\simeq 403\,G$, and $\Sigma/|\Delta| \simeq 0.054$. }
		\label{fig1b}
	\end{center}
\end{figure}

In the case of atomic condensates with a magnetic tunable resonance,  $\Sigma$ is usually negligible in the range of scattering lengths  experimentally accessible, corresponding to  weak inter-atom interactions with $0< a/\lambda_{T_c} \ll 0.01$. This is because it is difficult to stabilize the condensate for larger  $a$. Indeed, $\Sigma$  is commonly set to zero, so that $\tilde a$ is real and the scattering length is given by the well known  formula

\begin{equation}\label{scattering length 2}
a = a_{bg} \left(1-\frac{\Delta}{B-B_0}\right) \, .
\end{equation}
The values of $a_{bg}$ and $\Delta$ can be measured experimentally. For instance, one has $\Delta \simeq -52 \, G$, $a_{bg}\simeq -29\, a_0$, where $a_0$ is the Bohr radius, for the $B_0\simeq 403\,G$ resonance of $^{39}K$ \cite{d'errico}. The corresponding $a$ is plotted in Fig.\ref{fig1b} (dashed curve) in units of $\lambda_{T_c}\simeq 5.7 \times 10^{-7}m$.

Eq. (\ref{scattering length 2}) can be used to infer the value of the scattering length from the knowledge of the external magnetic field $B$. However, although this expression captures most of the features of the Feshbach resonance,  it is not accurate at  $B_0$, where it gives an infinite $a$, which is of course not physically meaningful. Moreover, in the strong interacting range $a/\lambda_{T_c} \gtrsim 0.01$ explored in \cite{condensatePRL} for the $403\,G$ Feshbach resonance of $^{39}K$ condensates, inelastic scattering processes become important, which implies that $\Sigma$ cannot be neglected. 


Thus, one has to retain $\Sigma$ in order to interpret the measurements of $\Delta T_c/T^0_c$ in \cite{condensatePRL}, so that the correct expression for the scattering length becomes

\begin{equation}\label{scattering length 3}
		a_p = a_{bg} \left(1-\frac{\left(B-B_0\right)/\Delta}{\left(\left(B-B_0\right)/\Delta\right)^2+\left(\Sigma/\Delta\right)^2}\right)\, .
\end{equation}
This function is plotted in fig.\ref{fig1b} (solid line) in unit of $\lambda_{T_c}\simeq 5.7 \times 10^{-7}m$ for the $403\,G$  Feshbach resonance of $^{39}K$, assuming  $\Sigma/|\Delta| \simeq 0.054$.
Indeed, while (\ref{scattering length 2}) diverges when $B$ approaches the resonance at $B_0$, (\ref{scattering length 3}) remains finite and it reaches its maximum   value at
\begin{equation}
		a^M_p = a_{bg}\left(1-\Delta/2\Sigma\right)\qquad \text{for} \qquad B=B_0-\Sigma \, .
\end{equation}
This means that, if (\ref{scattering length 2}) is used to infer the value of  the scattering length measuring $B$, $a$  is overestimated with respect to its real value $a_p$. 
This fact, causes the disagreement between the observed shift in $T_c$ and mean-field predictions, that has been reported in \cite{condensatePRL}.

\begin{figure}	\begin{center}
		\includegraphics[height=5cm]{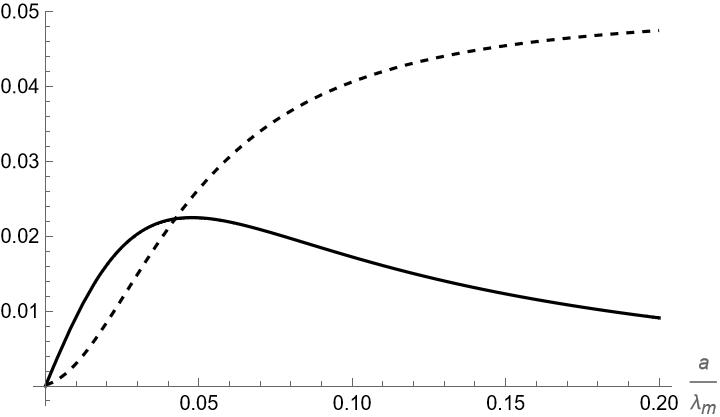}\qquad	\caption{We plot $a_p/\lambda_{T_c}$ and $b_p/\lambda_{T_c}$ against $a/\lambda_{T_c}$ as given by (\ref{scattering length 4},\ref{collisional loss b}) for $\lambda_{T_c}\simeq 5.7 \times 10^{-7}m$,   and $\Sigma/|\Delta| \simeq 0.054$. }
		\label{fig2b}
	\end{center}
\end{figure}

\begin{figure}	\begin{center}
		\includegraphics[height=5cm]{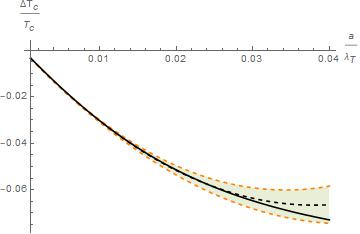}\qquad	\caption{We plot ${\Delta T_c}/{T^0_c}$ in Eq. (\ref{DT/T Final 2}) as a function of $a/\lambda_{T_c}$ (solid line) in the range of values explored in \cite{condensatePRL}. The   dashed curve is the best fit in (\ref{deltaTsuTsecondorder}), while the shadowed region is the $1\sigma$ contour.}
		\label{fig3}		
	\end{center}
\end{figure}

As we want to clean the distortion due to the overestimation of the scattering length, and compare the correct result with data, we can express the physical scattering length $a_p$ in terms of  its overestimated value $a$ as

\begin{equation}\label{scattering length 4}
a_p(a) = a_{bg} \left(1-\frac{\left(1- a/a_{bg}\right)^{-1}}{\left(1-a/a_{bg}\right)^{-2}+\left(\Sigma/\Delta\right)^2}\right),
\end{equation}
and replace $a\rightarrow a_p(a)$ in Eq. (\ref{DT/T Final}). The function (\ref{scattering length 4}) is plotted in fig.\ref{fig2b} (solid line) in unit of $\lambda_{T_c}\simeq 5.7 \times 10^{-7}m$ against $a/\lambda_{T_c}$  for $\Sigma/|\Delta|\simeq 0.054$ in the range \mbox{$0\leq a/\lambda_{T_c} \lesssim 0.2$.}

From Fig.\ref{fig1} it is evident that the measured temperature shift $\Delta T/T$ converges asymptotically to a constant value ($\sim -0.065$) for $a/\lambda_{T_c}\gtrsim 0.02$. Indeed, the data can be reproduced correctly imposing  that  $a^M_p/\lambda_{T_c}\sim 0.02$, corresponding to $a/\lambda_{T_c} \simeq 0.04$. In facts, $a_p(a)$ is approximately constant around its maximum, so that the corrected temperature shift will be nearly flat around $a/\lambda_{T_c} \sim 0.04$. 

Indeed, we set $a^M_p/\lambda_{T_c}= 0.023$, which gives $\Sigma/|\Delta|\simeq 0.054$, and then we replace $a\rightarrow a_p(a)$ in (\ref{DT/T Final}), obtaining the  rescaled temperature shift

\be\label{DT/T Final 2}
\frac{\Delta T_c}{T^0_c} = \left(\frac{I(0)}{I(a_p(a)/\lambda_{T_c})}\right)^{1/3}-1\, .
\ee
This function is plotted in Fig.\ref{fig3} for the $403 \, G$ resonance of the $^{39}K$ condensate. The temperature shift (\ref{DT/T Final 2}) stays always in the shadowed region, corresponding to the $1\,\sigma$ contour of (\ref{deltaTsuTsecondorder}), showing a quite perfect agreement between the mean-field prediction (\ref{DT/T Final 2}) and data.

It is probably worth  stressing that this conclusion does not rely on the expression (\ref{DT/T Final}), while it is based on the correct evaluation of the scattering length  given in (\ref{scattering length 4}) in terms  of the Feshbach magnetic field $B$. In facts, one might replace $a\rightarrow a_p(a)$ in the polynomial expression $\Delta T_c/T^0_c = b_1 (a/\lambda_{T_c})+ b_2 (a/\lambda_{T_c})^2$ with  $b_1 \simeq -3.42\ldots$ and $b_2 \simeq 12- 18$ as in \cite{condensatePRL2,briscese2}, obtaining a temperature shift compatible with (\ref{deltaTsuTsecondorder}) at $1\, \sigma$ level.

For completeness, we analyze the behavior of the collisional loss $b_p$ given by

\begin{equation}\label{collisional loss b}
	b_p = a_{bg} \frac{\Sigma/\Delta}{\left(\left(B-B_0\right)/\Delta\right)^2+\left(\Sigma/\Delta\right)^2}\, ,
\end{equation}
which has its maximum $b^M_p=a_{bg} \Delta/\Sigma$ at the resonance. In fig.\ref{fig1b} we plot the ratio $b_p/\lambda_{T_c}$ (dotted-dashed curve) against $B$. In fig.\ref{fig2b} we plot  the  same function (dashed curve) against $a/\lambda_{T_c}$, with its asymptotic limit $b^M_p/\lambda_{T_c}$ for $a\rightarrow \infty$, corresponding to the resonance.
We emphasize that the knowledge of $b_p$ allows to estimate the inelastic  rate coefficient by means of (\ref{kloss}), giving   $K_{loss} \simeq (1-5)\times 10^{-10} cm^3/s$ for $0.01\lesssim a\lambda_{T_c}\lesssim 0.04$.


In conclusion, we have shown that, taking into account the detailed form of the Feshbach resonance, it is possible to solve the long standing tension between the predicted temperature shift $\Delta T_c/T^0_c$ in the mean-field and semi-classical approximations, and the experiments performed with $^{39}K$ in the strong interacting regime. Such tension was due to an overestimation of the s-wave scattering length  due to the omission of loss effects encoded in the parameter $\gamma$. When such effects are taken into account, one finds a perfect agreement between mean-field predictions and experiments.

\vskip.1cm

\textbf{Acknowledgements}: The author is grateful to A. R. P. Lima, A. Pelster, and A. Cherny for useful discussions during the early stages of this work. This work was started while the author was working in the Department of Mathematics of the University of Padova, Italy.

\end{document}